\documentclass[twocolumn]{bmcart}

\usepackage{amsthm,amsmath,amssymb}
\usepackage{mathptmx}
\usepackage[english]{babel}
\usepackage[T1]{fontenc}
\usepackage[utf8]{inputenc} 
\usepackage{graphicx}
\usepackage{here}

\startlocaldefs
\endlocaldefs

\begin{document}

\begin{frontmatter}

\begin{fmbox}
\dochead{Research}


\title{Clinical Validation of two Surface Imaging Systems for Patient Positioning in Percutaneous Radiotherapy}


\author[
   addressref={UKE},                   
   corref={UKE},                       
   noteref={n1},                        
   email={Manfred.Wiencierz@uk-essen.de}   
]{\inits{MW}\fnm{Manfred} \snm{Wiencierz}}
\author[
   addressref={UKE},
   noteref={n1},                        
]{\inits{KK}\fnm{Kathrin} \snm{Kruppa}}
\author[
   addressref={UKE},
]{\inits{LL}\fnm{Lutz} \snm{L\"{u}demann}}


\address[id=UKE]{
  \orgname{Radiation and Tumor Clinic, Universit\"{a}tsklinikum Essen}, 
  \street{Hufelandstr. 55},                     %
  \postcode{45147}                              
  \city{Essen},                              
  \cny{Germany}                                    
}


\begin{artnotes}
\note[id=n1]{Equal contributor} 
\end{artnotes}



\begin{abstractbox}

\begin{abstract} 
\parttitle{Background} 
Precise patient positioning and thus precise positioning of the planning target volume (PTV) is a prerequisite for effective treatment in percutaneous radiation therapy. Conventional imaging modalities used to ensure exact positioning for treatment typically involve additional radiation exposure of the patient. A different approach to patient alignment without additional exposure is provided by optical 3D surface scanning and registration systems.

\parttitle{Methods} 
We investigated two systems (\textsc{Catalyst} and \textsc{AlignRT}), which allowed us to generate and validate surface images of 50 patients undergoing irradiation on a Tomotherapy system. We compared the positions proposed by these two surface imaging systems (SIS) with the actual adjustments in patient positions made on the basis of Megavoltage CT scans. With the two SIS, the surface of every single patient was repeatedly recorded throughout treatment. In addition, we performed a systematic analysis of both systems using a body phantom.

\parttitle{Results} 
At least 10 surface images were available from each patient included in the analysis. The large number of patients allowed us to subdivide the study population into three groups by body region treated: \textit{head \& neck}, \textit{chest}, and \textit{pelvic}. The results for each of these three body regions were separately analyzed. Thus, the focus of our investigation is on body regions registered with the two systems rather than tumor entities. For \textsc{AlignRT}, for instance, we found an improvement in positioning for the \textit{Pelvic} region of 0.9\,mm in 75\,\% and 6.4\,mm in 95\,\% of the collected data in comparison to conventional marker alignment.

\parttitle{Conclusions} 
The two systems were most accurate in the \textit{head \& neck} region, where conventional mask alignment is already sufficiently accurate. The other two body regions, \textit{pelvic} and \textit{chest}, might benefit from positioning by SIS. But for both systems there is still a need for improvement to reduce statistical variation.

\end{abstract}




\end{abstractbox}
\end{fmbox}

\end{frontmatter}



%
%
%
\section{Introduction}\label{Section: Introduction}
\noindent Optimal delivery of radiotherapy crucially depends on accurate and reproducible positioning of the patient over several fractionated irradiations. The margins around the clinical target volume (CTV) define the extent of the planning target volume (PTV). The margins should take all geometric variations and inaccuracies into account, especially patient positioning uncertainties \cite {2000:vanHerk}. External patient alignment and internal organ motion with respect to healthy tissue both contribute to the total positioning uncertainty. The use of high-precision techniques such as intensity-modulated radiotherapy (IMRT), volume-modulated arc therapy (VMAT), and helical Tomotherapy allow precise dose prescription and planning. Such precise irradiation techniques place high demands on the reproducibility of patient positioning for serial irradiations.

Proper patient positioning can be verified by a number of methods of varying sophistication. These methods may include the use of external markers and diverse MeV and keV x-ray systems with planar or computerized volumetric image acquisition. After implant of radio-opaque seeds planar as well as computerized image acquisition systems are able to verify both, patient setup and target structure. The daily or frequent use of these systems, however, leads to an additional radiation exposure of normal tissue. The typical treatment of cancer patients consists of 25 to 40 radiotherapy fractions. This means that the extra radiation exposure resulting from the use of these positioning systems is no longer negligible and becomes more of a concern the more irradiations a patient receives.

Positioning patients without harming intact tissue is the rationale for the development of new approaches not requiring ionizing radiation, for example, ultrasound systems \cite{2003:Trichter} or optical surface matching systems \cite{2000:Li,2003:Ploeger,2006:Li,2008:Brahme,2009:Krengli,2010:Peng}. The approach of \textsc{AlignRT} \cite{2009:Krengli,2010:Peng} and \textsc{Catalyst} use a three-dimen\-sional optical scan of the body surface for verification of the patient's position. This surface scan will be compared with the optimal (e.g., from a reference scan or a surface reconstruction from the planning CT scan) position. Deviations in all dimensions are calculated through body surface comparison but not for the target volume. The present study evaluates the accuracy of the calculated deviations. Both surface imaging systems (SIS) were mounted on a helical Tomotherapy radiation system for simultaneous use in the same patients. The systems use different techniques for image acquisition and registration.
%
%
%
%
\section{Surface Imaging Systems (SIS)}\label{Section: SIS}
\noindent The two SIS compared in this study are mounted on the To\-mo\-thera\-py irradiation system (\textsc{Accuray}, Madision, USA). The Tomo\-therapy device can acquire a CT scan of the patient with a de-tuned radiation energy of 3.5\,MeV \cite{2006:Mackie}, abbreviated in the following as MVCT. An MVCT of a patient positioned for treatment can be compared with the planning CT scan to directly match the patient's internal contours. The Tomotherapy device comprises a High-Performance Couch, which can be moved with an accuracy and precision in submillimeter range.
\begin{figure}[h]
  \includegraphics[width=0.45\textwidth]{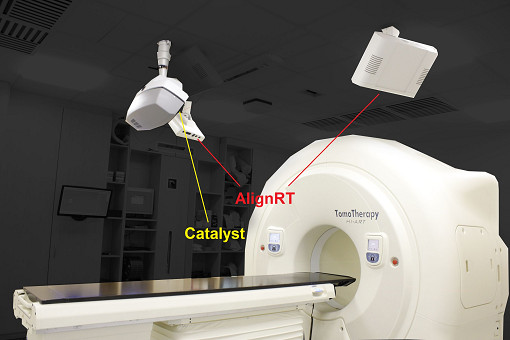}
  \caption{\csentence{- Setup of the SIS in the Tomotherapy treatment room.}
       The two SIS are attached to the ceiling of the room. The Catalyst scanner is mounted directly in front of the Tomotherapy gantry. The two scanners of the AlignRT are at the ceiling to the right and left of the couch.}
  \end{figure}
\subsection{Concept}\label{Section: SIS-Concept(sub)}
\noindent The basic approach of all SIS is the generation of a 3D patient surface using an optical scanner unit. The three-dimensional coordinates of each image point are determined by triangulation \cite{2001:Wiora}. The 3D surface can be compared with a reference surface. For patient treatment in percutaneous radiotherapy the reference surface can be currently the external body contour of the planning CT scan (external CT reference) \cite{2000:Doessel}, generated by a planning system, or a recorded image of the SIS (internal optical reference). The purpose of comparing the 3D patient surface with the reference surface (known as registration) is to calculate the geometric shifts from the recorded surface to the reference. Hence, it is important for users to know which type of reference will ensure optimal registration. After registration the SIS computes geometrical adjustments to align the recorded 3D surface to the reference. In general, these shifts include translational (lateral $\Delta\textrm{x}$, longitudinal $\Delta\textrm{y}$ and vertical $\Delta\textrm{z}$) and rotational (angles $\phi$ around x-axis, $\vartheta$  around y-axis and $\gamma$ around z-axis) values.
\subsection{AlignRT}\label{Section: SIS-Align RT(sub)}
\noindent The AlignRT is marketed by \textsc{VisionRT} (London, UK) and, in the version installed at our facility, basically consists of two scanner units. Figure 1 shows the setup of the scanners (denoted by red lines) in the Tomotherapy treatment room. They are mounted at the ceiling to the right and left of the Tomotherapy couch. Each of the two scanners includes several components (stereoscopic camera, speckle projector, texture camera, light flash) \cite{2010:VisionRT}. In combination with the light flash, the texture camera captures a gray-scale image of the patient. This gray-scale image only provides visual information for the relevant scan region \cite{2005:Bert}. The 3D surface is directly captured by the stereoscopic camera over (passive) triangulation \cite{2007:Hering,2005:Schreer}. For improved triangulation, a pseudorandom gray-scale pattern is projected (speckle projector) onto the patient to reduce incorrect reflections \cite{2012:Willoubhby}. The field of view from a single scanner unit does not fully capture the patient surface. Due to the geometric constellation \cite{2009:Krengli} the patient himself conceals the missing surface. For this reason, the second scanner unit captures the patient surface from another angle. The two surfaces from both scanner units are merged by software to produce a nearly complete 3D surface of the patient. Figure 2 shows an example of a 3D surface recorded by AlignRT. This surface can then be used for registration. The AlignRT system uses a rigid registration algorithm, which means that the patient body is regarded as a rigid system.
\subsection{Catalyst}\label{Section: SIS-Catalyst(sub)}
\noindent The Catalyst system, from \textsc{C-Rad} (Uppsala, Sweden), consist only of a single scanner unit. Figure 1 shows the setup of the scanner in the Tomotherapy treatment room (denoted by a yellow line). It is mounted at the ceiling directly in front of the Tomotherapy gantry. The scanner includes a single CMOS camera and a structured light projector. The principle of 3D surface capturing (detailed description in \cite{2007:Ishii}) differs from AlignRT. By the projection of a stripe pattern onto the patient's surface the triangulation is here considered to be active. The dispersion of the projected pattern on the patient is detected by the camera. Figure 2 shows an example of a 3D surface generated by Catalyst. This surface can then be used for registration. The Catalyst system uses a non-rigid registration algorithm. This non-rigid algorithm can detect local deformations in the captured surface. This extra information maximizes the region of overlap \cite{2008:Li}. An additional feature is that the detected deformation is directly projected onto the patient's surface and can assist in adjusting the position.
  \begin{figure}[t]
  \includegraphics[width=0.45\textwidth]{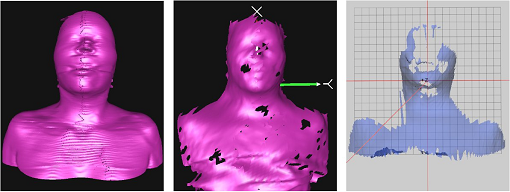}
  \caption{\csentence{- Examples of 3D surfaces.}
       All surfaces were generated from the same patient. The surface shown on the left was generated by contouring a CT scan. The surface in the center was generated by Align RT and the surface on the right by Catalyst.}
  \end{figure}
%
%
%
\section{Phantom Measurement}
\noindent  The phantom measurements were performed using an Alderson Radiation Therapy phantom (\textsc{RSD}, Long Beach, USA). The Alderson phantom was precisely aligned on the couch. With well-defined displacements of the couch both SIS should validate these displacements. Due to the use of both reference modes (external CT and internal optical reference) this validation procedure was performed twice. These checks were intended to expose possible systematic errors in contouring, data export (between Tomotherapy planning system and SIS), and alignment.
\subsection{Systematic Procedure}
\noindent In detail, we used an MVCT to determine the exact position of the Alderson phantom in relation to the external CT reference. At this stage, we also captured the internal optical references from both systems to create a sustainable comparison. The well-defined couch displacements were applied separately for the \textit{x}- , \textit{y}- and \textit{z}-dimension. The range of displacements in \textit{y}- and \textit{z}-dimension was $[-50\,\textrm{mm};50\,\textrm{mm}]$; whereat the couch was displaced in discrete steps (mm) of $y,z\in\{\pm50,\pm35,\pm25,\pm20,\pm15,$ $\pm10,\pm5,0\}$. Due to the couch limitations in \textit{x}-dimension, the displacement range was $[-25\,\textrm{mm}, 25\,\textrm{mm}]$, with equidistant steps of 5\,mm. At each of these couch displacements the surface of the Alderson phantom was captured and registered with the internal optical reference of both systems. The registration of these surfaces with the external CT reference was performed later, in the postprocessing mode.

The aberration of SIS registration arising from couch displacement is calculated as the difference of the proposed deviation of the SIS and the adjusted couch deviation. The results are presented for each dimension in \textit{x}, \textit{y} and \textit{z}. However, we are only interested in the absolute aberration and not in directional dependence in one dimension. With this approach, we can dispense an averaging of the results near zero in the statistical analysis. The distribution of the results received from the internal optical reference is denoted by $|\Delta C(\vec{r})|$ (Catalyst) and $|\Delta A(\vec{r})|$ (AlignRT) in capital letters, whereat $\vec{r}=(x,y,z)$. The notation for the results with use of the external CT reference is in lowercase letters to identify the SIS ($|\Delta c(\vec{r})|$ and $|\Delta a(\vec{r})|$).\newline

  \begin{figure*}
  \includegraphics[width=0.9\textwidth]{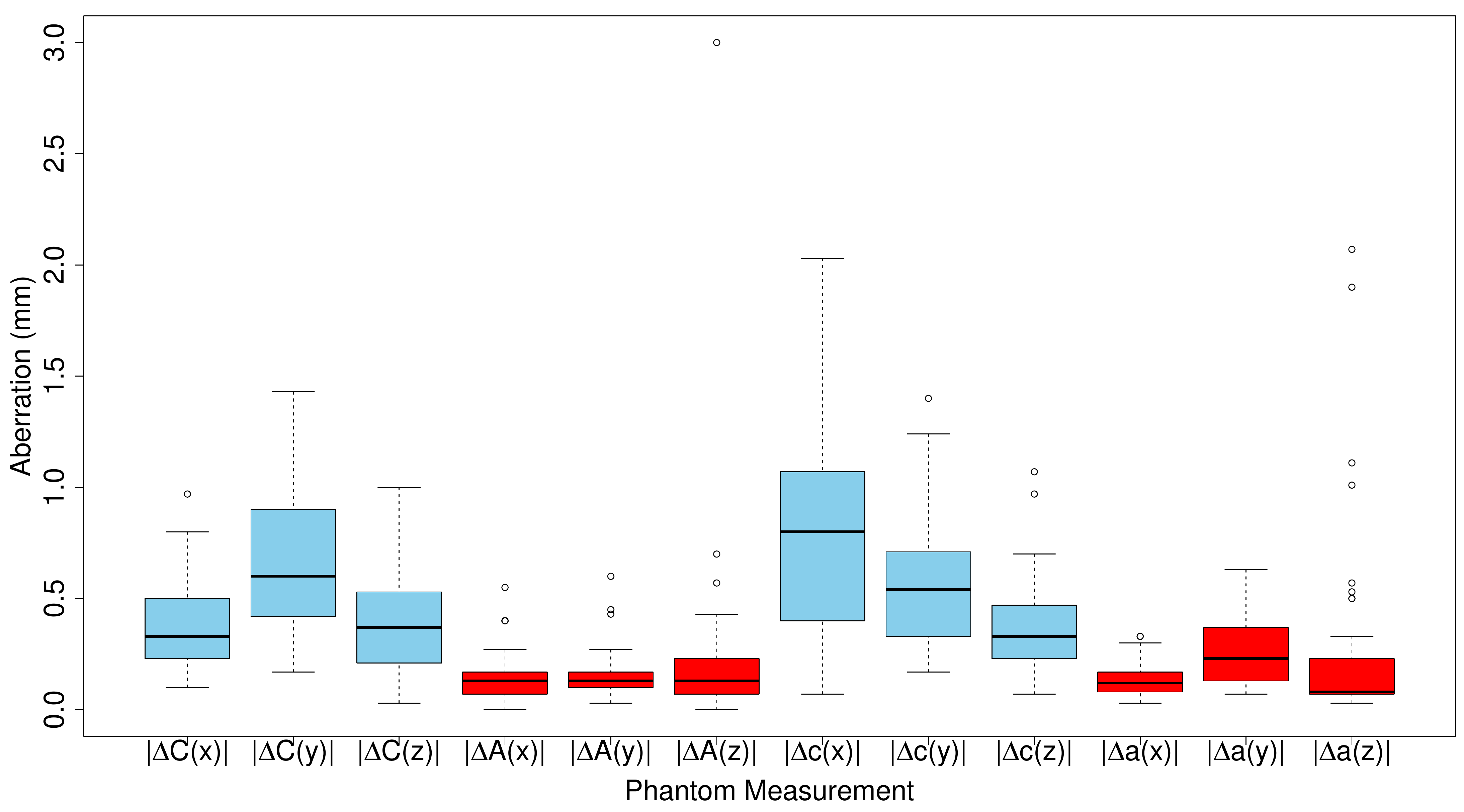}
  \caption{\csentence{- Boxplot of the distributions derived from phantom measurements.}
       The distributions obtained using a SIS image as reference are labeled with uppercase letters. The lowercase letters indicate distributions obtained using a reference image from the planning CT scan.}
    \end{figure*}

\noindent A look at Figure 2 illustrates the different surface recording coverage of the two SIS. For this reason, both SIS were analyzed by an angle coverage investigation. This was done by placing a cylinder-shaped phantom (\textsc{Gammex}, Middleton, USA) with equidistant markers on the curved surface on the couch. The curved surface was captured by both SIS for the angles $\phi$ and $\vartheta$. Using the markers on the phantom surface, we could estimate the angle coverage for surface capturing.
\begin{table}[b!]
\caption{Listings of the statistical percentiles - median \textsf{M}, \textsf{P$_{75}$}, \textsf{P$_{90}$} and \textsf{P$_{95}$} - of the value distributions. Distributions for use of an internal optical reference $|\Delta C|$ and $|\Delta A|$ are labeled with uppercase letters; the lowercase letters indicate distributions $|\Delta c|$ and $|\Delta a|$ obtained with use of a reference image from an external CT scan.}
      \begin{tabular}{ccccc}\hline
        Distribution	&	\textsf{M} (mm)	&	\textsf{P$_{75}$} (mm)	&	\textsf{P$_{90}$} (mm)	&	 \textsf{P$_{95}$} (mm)	\\	\hline
        $|\Delta C(x)|$	&	0.3	&	0.5	&	0.6	&	0.8	\\	
        $|\Delta C(y)|$	&	0.6	&	0.9	&	1.2	&	1.2	\\
        \vspace{1.5mm}$|\Delta C(z)|$	&	0.4	&	0.5	&	0.8	&	0.8	\\	
        $|\Delta A(x)|$	&	0.1	&	0.2	&	0.4	&	0.4	\\	
        $|\Delta A(y)|$	&	0.1	&	0.2	&	0.3	&	0.4	\\
        \vspace{1.5mm}$|\Delta A(z)|$	&	0.1	&	0.2	&	0.4	&	0.6	\\	
        $|\Delta c(x)|$	&	0.8	&	1.1	&	1.3	&	1.4	\\	
        $|\Delta c(y)|$	&	0.5	&	0.7	&	1.1	&	1.2	\\
        \vspace{1.5mm}$|\Delta c(z)|$	&	0.3	&	0.5	&	0.6	&	0.7	\\	
        $|\Delta a(x)|$	&	0.1	&	0.2	&	0.3	&	0.3	\\	
        $|\Delta a(y)|$	&	0.2	&	0.4	&	0.6	&	0.6	\\	
        $|\Delta a(z)|$	&	0.1	&	0.2	&	0.6	&	1.1	\\	\hline
      \end{tabular}
\end{table}
\subsection{Results}
\noindent Figure 3 presents boxplots of all distributions. Because we only considered absolute values, our results are not normally distributed. Hence, only the deviations above the median are significant. The corresponding statistical parameters - median \textsf{M}, the 75th percentile \textsf{P$_{75}$} (upper quartile in boxplot), the 90th percentile \textsf{P$_{90}$}, and the 95th percentile \textsf{P$_{95}$} - are presented in Table 1. For a feasible clinical applicability of the two SIS, the last two percentiles permit to estimate the maximum variation of the distributions without statistical outliers \cite{2000:vanHerk}.

The accuracy in positioning from all distributions is in median value \textsf{M} below 1\,mm. With the exception of $|\Delta C(y)|$, $|\Delta c(x)|$, $|\Delta c(y)|$ and $|\Delta a(z)|$ the accuracy in positioning is even in \textsf{P$_{95}$} below 1\,mm. The high value for $|\Delta a(z)|$ in \textsf{P$_{95}$} can, however, be explained by the larger number of outliers in this distribution, while the other percentiles exhibit moderate values. These results are in the same range as those reported in \cite{2005:Bert} and \cite{2007:Schoeffel}. Overall, our phantom measurements reveal lower values for AlignRT. The comparison of the two reference types did not reveal any true advantage of either of the two for positioning.\newline

\noindent The angle coverage investigation confirms our assumption that AlignRT offers extended coverage. For coverage by Catalyst we found values of $\phi_{C}\approx150\,^{\circ}$ and $\vartheta_{C}\approx120\,^{\circ}$, whereas AlignRT yielded values of $\phi_{A}\approx150\,^{\circ}$ and $\vartheta_{A}\approx220\,^{\circ}$. The extended coverage in the range of $\vartheta$ is definitely attributable to the use of a second camera in the AlignRT.
%
%
\section{Clinical Evaluation}
\noindent Both SIS were tested under clinical conditions. But the surface registration systems were not be used in the current study as medical products according to \S\,3 Abs. 1 \textsf{Medizinproduktegesetz – MPG}. However, the proposed results of SIS registration were not used to actually modify patient positioning for radiotherapy. Informed consent was therefore not required for this retrospective study. In detail, patients were setup in align position using signed markers. At this stage, both SIS captured the 3D surface of the patient and provide proposal shifts by registration with the external CT reference to adjust the patient in \textit{correct} position. The software of both SIS enables registration of the captured 3D surface with the internal optical reference in postprocessing mode. Therefore, both types of reference were used for registration. The patient align position is checked by an MVCT scan. Patient adjustment $\mu(\vec{r})$ from align to treatment position based on MVCT is performed by a radiation therapist with more than five years of experience. These adjustment results, $\mu(\vec{r})$, were recorded for every treatment.

The notations for the distributions in the clinical evaluation, where an internal optical image serves as reverence, differs from those used for the phantom experiments. The following expressions are defined:
\begin{align}
\Delta A(\vec{r})=\left|A(\vec{r})-\mu(\vec{r})\right|,\\
\Delta C(\vec{r})=\left|C(\vec{r})-\mu(\vec{r})\right|.
\end{align}
$C(\vec{r})$ and $A(\vec{r})$ are the shifts proposed by the SIS on the basis of the internal optical reference. The shifts $\mu(\vec{r})$ are derived from the MVCT scan and were applied to make adjustments in patient position when moving the patient from align to treatment position. Hence, $\Delta C(\vec{r})$ and $\Delta A(\vec{r})$ are the final aberrations calculated for the two SIS with use of an internal optical image as reference.

Unfortunately, patient position and shape differ between the diagnostic CT scanner and treatment unit due to different equipment and patient comfort. This situation, however, does not affect radiation treatment since, with use of MVCT scans only the internal structures were matched with the CT scan to ensure the ideal patient position for treatment. To estimate the patient setup error between the images acquired in the CT (used as external reference) and the actual patient position in the Tomotherapy, we averaged the first three registration results and subtracted this average from subsequent registrations. This approach means that the shifts proposed by the SIS must first be corrected by
\begin{equation}
\overline{a_{3}(\vec{r})}=\frac{1}{3}\cdot\sum\limits_{i=1}^{3}a_{i}(\vec{r}), \quad
\overline{c_{3}(\vec{r})}=\frac{1}{3}\cdot\sum\limits_{i=1}^{3}c_{i}(\vec{r}).
\end{equation}
The correction values, $\overline{c_{3}(\vec{r})}$ for Catalyst and $\overline{a_{3}(\vec{r})}$ for AlignRT, are determined individually for each patient. Finally, aberrations were calculated using the following expressions:
\begin{align}
\Delta a(\vec{r})=\left|a(\vec{r})- \overline{a_{3}(\vec{r})}-\mu(\vec{r})\right|,\\
\Delta c(\vec{r})=\left|c(\vec{r})- \overline{c_{3}(\vec{r})}-\mu(\vec{r})\right|.
\end{align}
As in the previous section, the aberrations in registration from external CT reference $\Delta c(\vec{r})$ for Catalyst and $\Delta a(\vec{r})$ for AlignRT are denoted by lowercase letters.
\begin{table}[h!]
\caption{Listings of the statistical percentiles - median \textsf{M}, \textsf{P$_{75}$}, \textsf{P$_{90}$} and \textsf{P$_{95}$} - of the value distributions. Distributions for use of an internal optical reference $\Delta C$ and $\Delta A$ are labeled with uppercase letters; the lowercase letters indicate distributions $\Delta c$ and $\Delta a$ obtained with use of a reference image from an external CT scan. The three distributions of $\mu$ describe the mismatch between the align position, which is indicated by markers, and the treatment position.}
      \begin{tabular}{ccccc}\hline
        Distribution	&	\textsf{M} (mm)	&	\textsf{P$_{75}$} (mm)	&	\textsf{P$_{90}$} (mm)	&	 \textsf{P$_{95}$} (mm)	\\	\hline
        $\Delta C(x)$	&	1.1	&	2.0	&	3.0	&	3.5	\\	
        $\Delta C(y)$	&	1.5	&	3.0	&	4.7	&	6.3	\\	
        \vspace{1.5mm}$\Delta C(z)$	&	1.0	&	1.7	&	2.3	&	2.7	\\	
        $\Delta A(x)$	&	1.0	&	1.9	&	2.8	&	3.9	\\	
        $\Delta A(y)$	&	1.4	&	2.1	&	3.3	&	4.0	\\	
        \vspace{1.5mm}$\Delta A(z)$	&	2.1	&	2.7	&	4.7	&	6.3	\\	
        $\Delta c(x)$	&	1.9	&	3.8	&	6.1	&	8.1	\\	
        $\Delta c(y)$	&	2.1	&	3.5	&	5.6	&	7.3	\\	
        \vspace{1.5mm}$\Delta c(z)$	&	1.5	&	3.4	&	6.7	&	7.6	\\	
        $\Delta a(x)$	&	0.9	&	1.8	&	2.7	&	3.7	\\	
        $\Delta a(y)$	&	1.4	&	2.3	&	3.2	&	3.6	\\	
        \vspace{1.5mm}$\Delta a(z)$	&	0.9	&	1.9	&	2.9	&	3.5	\\	
        $\mu(x)$	    &	1.3	&	2.4	&	3.4	&	4.2	\\	
        $\mu(y)$	    &	1.2	&	2.0	&	2.9	&	3.8	\\	
        $\mu(z)$	    &	0.8	&	1.5	&	2.0	&	2.6	\\	\hline
      \end{tabular}
\end{table}
   \begin{figure*}
   \includegraphics[width=0.9\textwidth]{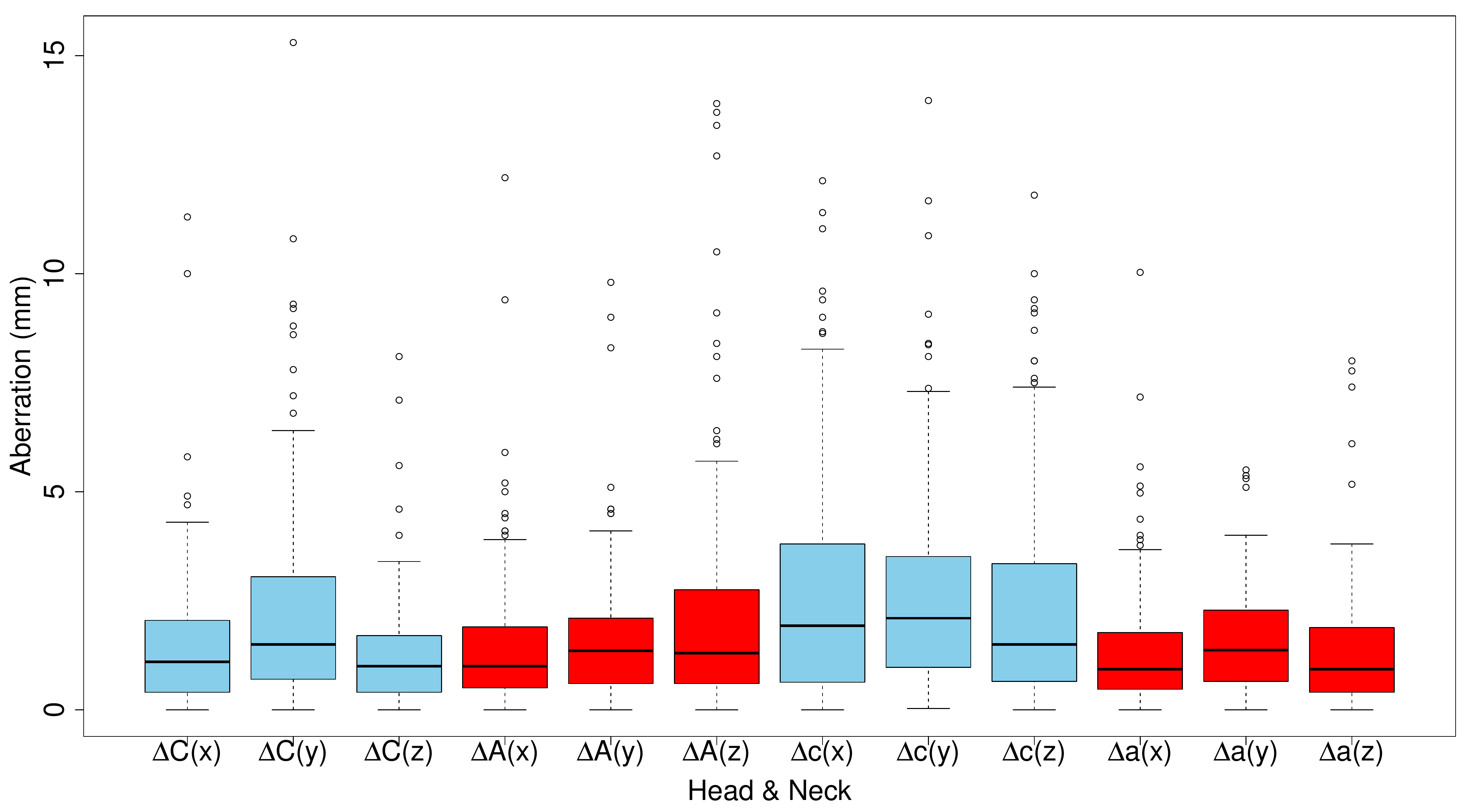}
   \caption{\csentence{- Boxplot of the distributions obtained for patient registration in the \textit{head \& neck} body region.}
       The distributions obtained using a SIS image as reference are labeled with uppercase letters. The lowercase letters indicate distributions obtained using a reference image from the planning CT scan.}
    \end{figure*}
\subsection{Patient Selection}
\noindent During the evaluation period, 50 patients with different tumor entities were treated. Thereby diverse tumor entities are arranged in the same body region. Therefore, we decided to group tumors by three body regions \textit{head \& neck}, \textit{pelvic} and \textit{chest}. For a meaningful analysis every patient had to satisfy the following criteria:
\begin{itemize}
\item The amount of captured surface must be sufficient for meaningful registration. The area of surface captured depends on the camera settings (individually adjusted for every patient) and the amount of uncovered patient skin or surface. Especially in the \textit{Pelvic} region not all patients are willing to do without coverings.
\item All patients positioned using an aircast vacuum mattress were excluded. Both SIS cannot distinguish between patient surface and the surface of the mattress. This selection criterion again excluded some patients with tumors of the \textit{pelvic} region.
\item To obtain enough data for statistical analysis and further evaluation, at least ten usable surface recordings had to be available from a patient.
\end{itemize}
\noindent After this selection procedure, 19 patients remained for evaluation. They were distributed as follows among the three body regions. The listing provides additional information on the tumors treated, type of immobilization, and other specific features.
\begin{description}
\item[Head \& Neck:] Ten patients were included for evaluation. They had tumors in the brain, the nasopharynx, the oropharynx, and on the skin. All patients were fixated on the couch with individually prepared head masks (\textsc{Civco} Posicast, Kalona, USA). These masks cover large parts of the interesting surface for registration. Both SIS can only capture the surface of the mask and not that of the patient's head.
\item[Pelvic:] Five patients were available for evaluation. Tumors were located in the prostate, the rectum, and the pelvic bone. For reproducible positioning the legs of all patients were aligned on a knee cushion and the feet were fixed. These measures are taken to minimize undesired leg rotations during treatment and ensure a consistent position of the pelvic region.
\item[Chest:] Four patients with chest tumors met the criteria for inclusion. They had tumors of the lung or esophagus. As in the \textit{head \& neck} group, all patients were fixated with a head mask. However, in this group, the mask does not cover the body surface in the area of the PTV.
\end{description}
\subsection{Results}
\noindent The results of the clinical evaluation are presented separately for each of the three body regions investigated. The results are displayed in boxplots in Figs. 4 -- 6. The statistical parameters are presented in Tables 2 -- 4. Unlike the phantom measurements, the clinical adjustments made $\mu{(\vec{r})}$ are also listed in the tables. With the help of $\mu{(\vec{r})}$ it is possible to compare the abberations of the SIS with the precision of the conventional marker alignment.
\subsubsection*{Head \& Neck}
\noindent Figure 4 and Table 2 present the results of the distributions in the body region \textit{head \& neck}. For registration with the external CT reference, all distributions from Catalyst ($\Delta c(\vec{r})$) show above-average values in all percentiles. Furthermore, in comparison to the other distributions ($\Delta C(x)$ and $\Delta C(z)$),  $\Delta C(y)$ exhibits uncommonly high values in the \textsf{P$_{75}$}, \textsf{P$_{90}$}, and \textsf{P$_{95}$} percentiles. A similar abnormality is apparent for $\Delta A(z)$ to $\Delta A(x)$ and $\Delta A(y)$ in all percentiles. The most accurate results are provided by AlignRT when an external CT scan is used as reference. All distributions concerned $\Delta a(\vec{r})$ show an accuracy of \textsf{P$_{75}$} $\leq$ 2.3\,mm and \textsf{P$_{95}$} $\leq$ 3.7\,mm. The comparison of the most accurate results from $\Delta a(\vec{r})$ with the results of conventional marker alignment shows an almost similar accuracy in positioning for marker alignment. All distributions $\mu (\vec{r})$ show an accuracy of \textsf{P$_{75}$} $\leq$ 2.4\,mm and \textsf{P$_{95}$} $\leq$ 4.2\,mm.
\subsubsection*{Pelvic}
\noindent Figure 5 and Table 3 present the results of the distributions in the body region \textit{Pelvic}. Again, both longitudinal distributions of Catalyst $\Delta C(y)$ and $\Delta c(y)$ show conspicuously high values in all percentiles. Furthermore, the values of $\Delta C(x)$ in this body region are also unusually high in all percentiles. Otherwise, only the distribution $\Delta a(z)$ of AlignRT again exhibits high values in all percentiles compared with $\Delta a(x)$ and $\Delta a(y)$. In addition, the  $\Delta A(x)$ and $\Delta c(z)$ distributions offer high values for the \textsf{P$_{90}$} and \textsf{P$_{95}$} percentiles. These values can be explained by a large number of outliers in these distributions, while the \textsf{P$_{75}$} and \textsf{P$_{90}$} percentiles show moderate values. The most accurate results are again achieved with AlignRT using an external CT scan as reference. All distributions concerned $\Delta a(\vec{r})$ show an accuracy of \textsf{P$_{75}$} $\leq$ 5.9\,mm and \textsf{P$_{95}$} $\leq$ 9.8\,mm. In comparison to conventional marker alignment, $\Delta a(\vec{r})$ clearly improves positioning accuracy in \textsf{P$_{75}$} and \textsf{P$_{95}$}. All distributions using the body markers $\mu (\vec{r})$ show an accuracy of \textsf{P$_{75}$} $\leq$ 6.8\,mm and \textsf{P$_{95}$} $\leq$ 16.2\,mm. Similar values for positioning accuracy with conventional markers in prostate treatment were reported in \cite{2006:Bortfeld}.
\subsubsection*{Chest}
\noindent Figure 6 and Table 4 show the results of the distributions in the body region \textit{chest}. Only the two longitudinal distributions of Catalyst show prominently high values in this body region. These high values are found for $\Delta C(y)$ in \textsf{P$_{75}$}, \textsf{P$_{90}$} and \textsf{P$_{95}$} and for $\Delta c(y)$ in all percentiles. All other distributions offer moderate values in all percentiles. The most accurate results are again provided by AlignRT using an external CT Scan as reference. All distributions concerned $\Delta a(\vec{r})$ show an accuracy of \textsf{P$_{75}$} $\leq$ 3.2\,mm and \textsf{P$_{95}$} $\leq$ 6.1\,mm. Compared with conventional marker alignment, $\Delta a(\vec{r})$ improves positioning accuracy. All distributions $\mu (\vec{r})$ show an accuracy of \textsf{P$_{75}$} $\leq$ 5.0\,mm and \textsf{P$_{95}$} $\leq$ 8.1\,mm.\newline

\noindent All distributions for the three body regions investigated were statistically analyzed by regarding the p-value with the null hypothesis of randomly distributed values. Except for $\Delta A(z)$ in the \textit{head \& neck} region and $\Delta C(x)$ in the \textit{pelvic} region, all distributions exhibit p-values of less than or equal to 0.01. The probability of randomly collected values is therefore less than or equal to 1\,\%. These distributions can be considered to be statistically significant. The remaining distributions could not be assigned a p-value.
  \begin{figure*}
  \includegraphics[width=0.9\textwidth]{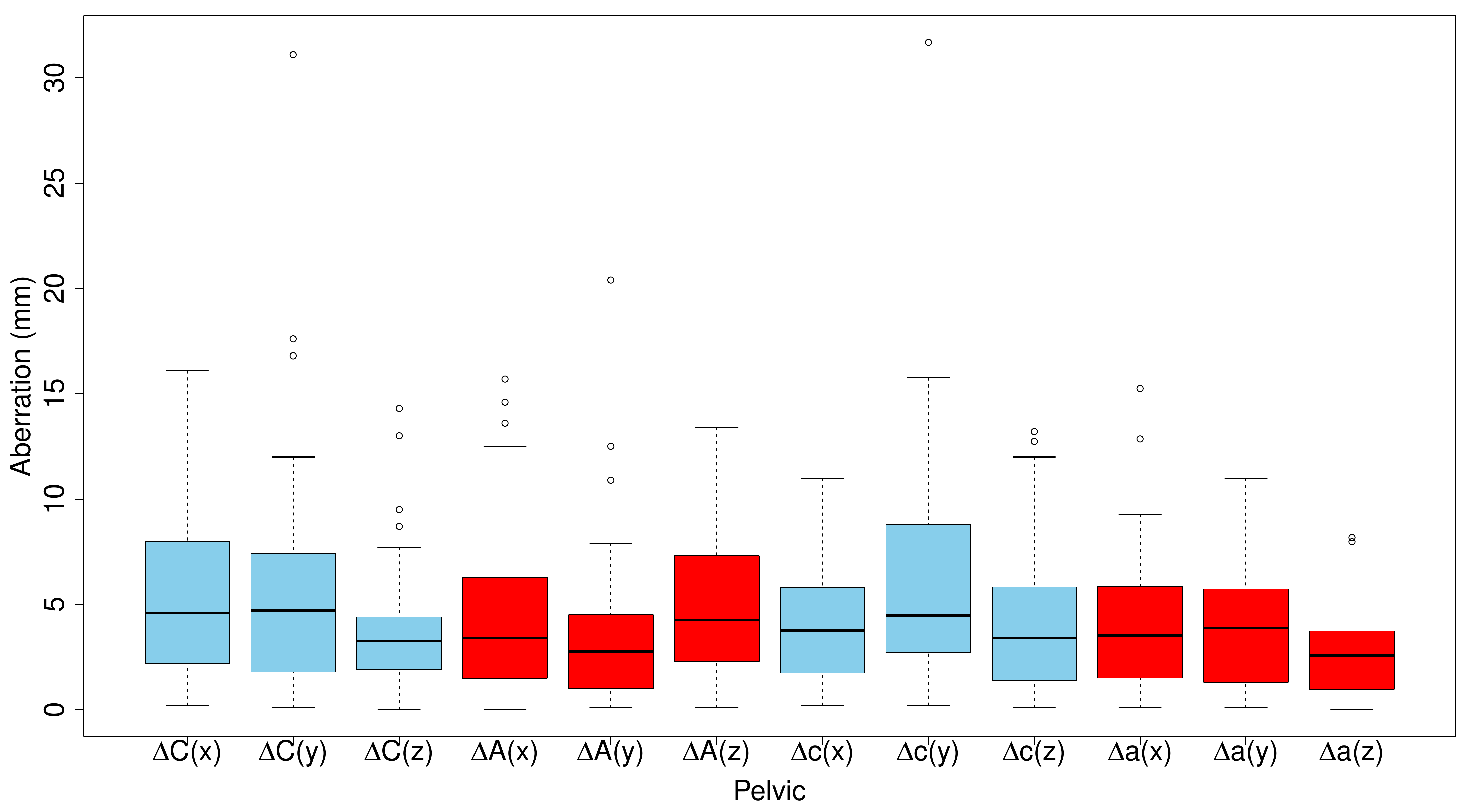}
    \caption{\csentence{- Boxplot of the distributions obtained for patient registration in the \textit{pelvic} body region.}
       The distributions obtained using a SIS image as reference are labeled with uppercase letters. The lowercase letters indicate distributions obtained using a reference image from the planning CT scan.}
   \end{figure*}
\begin{table}[b!]
\caption{See caption for Tab. 2.}
      \begin{tabular}{ccccc}\hline
        Distribution	&	\textsf{M} (mm)	&	\textsf{P$_{75}$} (mm)	&	\textsf{P$_{90}$} (mm)	&	 \textsf{P$_{95}$} (mm)	\\	\hline
        $\Delta C(x)$	&	4.6	&	8.0	&	11.5	&	12.1	\\	
        $\Delta C(y)$	&	4.7	&	7.3	&	10.8	&	11.9	\\	
        \vspace{1.5mm}$\Delta C(z)$	&	3.3	&	4.4	&	6.9	    &	8.1	    \\	
        $\Delta A(x)$	&	3.4	&	6.3	&	10.7	&	12.4	\\	
        $\Delta A(y)$	&	2.8	&	4.4	&	6.3	    &	7.5	    \\	
        \vspace{1.5mm}$\Delta A(z)$	&	4.3	&	7.3	&	10.1	&	11.3	\\	
        $\Delta c(x)$	&	3.8	&	5.8	&	6.6	    &	7.9	    \\	
        $\Delta c(y)$	&	4.5	&	8.8	&	12.4	&	15.1	\\	
        \vspace{1.5mm}$\Delta c(z)$	&	3.4	&	5.8	&	7.2	    &	11.5	\\	
        $\Delta a(x)$	&	3.5	&	5.9	&	7.3	    &	8.9	    \\	
        $\Delta a(y)$	&	3.9	&	5.7	&	8.6	    &	9.8	    \\	
        \vspace{1.5mm}$\Delta a(z)$	&	2.6	&	3.7	&	6.5	    &	7.5	    \\	
        $\mu(x)$	    &	4.0	&	6.8	&	12.6	&	16.2	\\	
        $\mu(y)$	    &	2.3	&	3.8	&	5.9    	&	6.6	    \\	
        $\mu(z)$	    &	1.9	&	3.2	&	5.5    	&	6.7	    \\	\hline
      \end{tabular}
\end{table}
%
%
\section{Discussion}
  \begin{figure*}
  \includegraphics[width=0.9\textwidth]{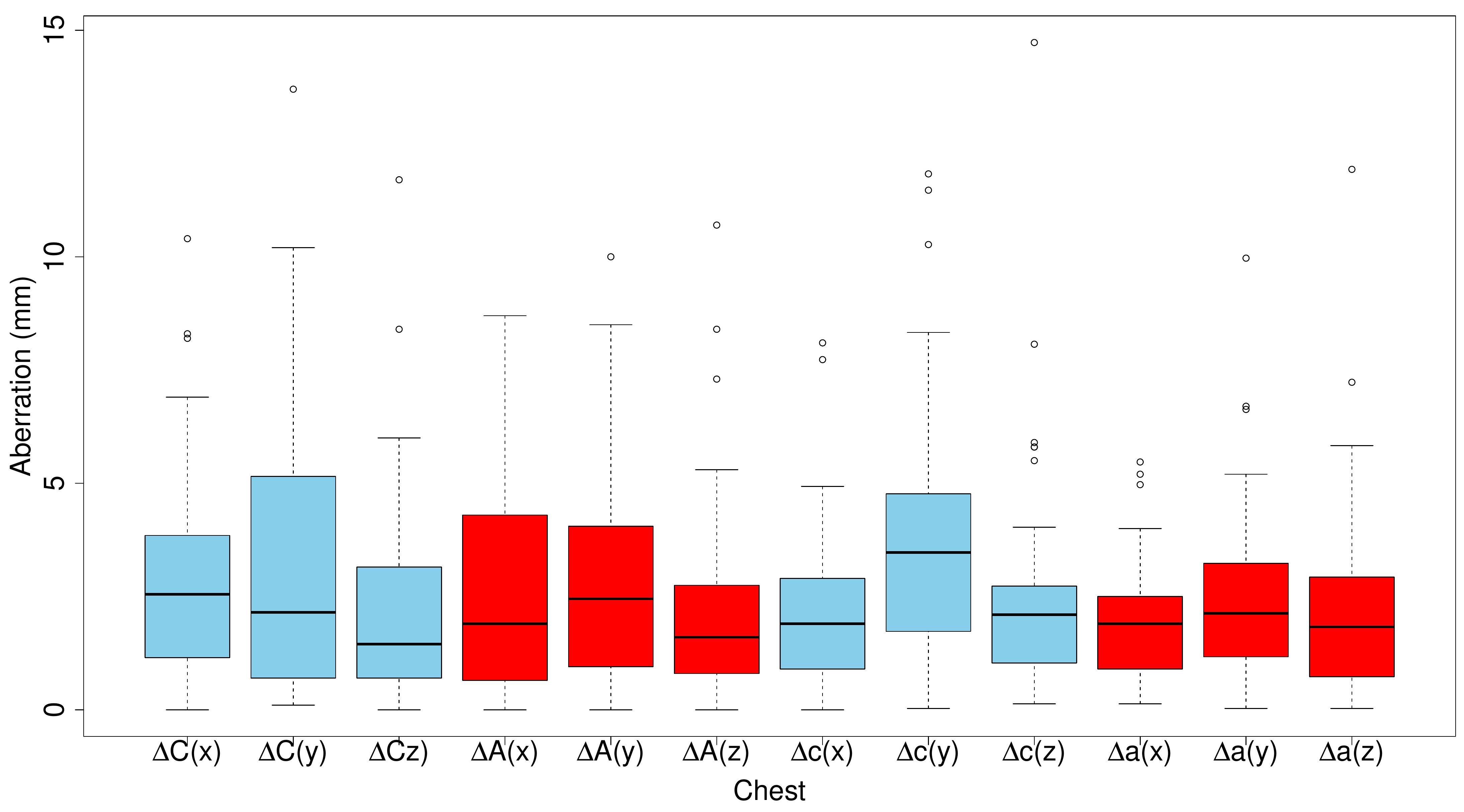}
  \caption{\csentence{- Boxplot of the distributions obtained for patient registration in the \textit{chest} body region.}
       The distributions obtained using a SIS image as reference are labeled with uppercase letters. The lowercase letters indicate distributions obtained using a reference image from the planning CT scan.}
  \end{figure*}
\begin{table}[b!]
\caption{See caption for Tab. 2.}
      \begin{tabular}{ccccc}\hline
        Distribution	&	\textsf{M} (mm)	&	\textsf{P$_{75}$} (mm)	&	\textsf{P$_{90}$} (mm)	&	 \textsf{P$_{95}$} (mm)	\\	\hline
        $\Delta C(x)$	&	2.6	&	3.8	&	5.7	&	7.0	\\	
        $\Delta C(y)$	&	2.2	&	5.0	&	7.3	&	8.7	\\	
        \vspace{1.5mm}$\Delta C(z)$	&	2.5	&	3.1	&	4.7	&	5.5	\\	
        $\Delta A(x)$	&	1.9	&	4.3	&	6.0	&	7.1	\\	
        $\Delta A(y)$	&	2.5	&	4.0	&	6.1	&	7.3	\\	
        \vspace{1.5mm}$\Delta A(z)$	&	1.6	&	2.7	&	4.9	&	5.4	\\	
        $\Delta c(x)$	&	1.9	&	2.9	&	4.4	&	4.9	\\	
        $\Delta c(y)$	&	3.5	&	4.8	&	7.3	&	9.5	\\	
        \vspace{1.5mm}$\Delta c(z)$	&	2.1	&	2.7	&	5.6	&	5.9	\\	
        $\Delta a(x)$	&	1.9	&	2.5	&	3.4	&	4.6	\\	
        $\Delta a(y)$	&	1.1	&	3.2	&	4.8	&	6.1	\\	
        \vspace{1.5mm}$\Delta a(z)$	&	1.8	&	2.9	&	4.7	&	5.8	\\	
        $\mu(x)$	    &	2.7	&	4.0	&	5.8	&	6.5	\\	
        $\mu(y)$	    &	2.8	&	5.0	&	7.0	&	8.1	\\	
        $\mu(z)$	    &	1.1	&	2.2	&	2.9	&	3.2	\\	\hline
      \end{tabular}
\end{table}
\noindent In both, the phantom measurements and the clinical evaluation, some distributions point considerable problems in positioning accuracy with both SIS investigated. Conspicuously high values were found for the four distributions $\Delta A(z)$, $\Delta C(x)$, $\Delta c(x)$ and $\Delta c(z)$ in at least one body region. Neither of the two SIS could not be insinuate a systematic inaccuracy in positioning in these distributions. For the distributions $\Delta C(y)$ and $\Delta c(y)$ of Catalyst, the situation is different. Here $\Delta C(y)$ and $\Delta c(y)$ provide overall high values in positioning accuracy. These results point to an inherent inaccuracy in longitudinal positioning. This systematic inaccuracy might result from coverage of a significantly smaller body surface area compared with AlignRT. It is likely that the longitudinal positioning inaccuracy affects the other results as well.

In all body regions, AlignRT in combination with an external CT scan as reference provides the most accurate results in positioning. Without the longitudinal error, the Catalyst might provide more precise positioning results using the same CT reference, due to the bigger surface coverage. In general, it is wise to choose the reference that covers the largest body surface area. However, use of the currently available external CT reference is more time consuming than use of the SIS internal reference, precluding its integration into clinical workflow. Additionally, the patient setup error has to be corrected first. In the present study, we used the average of the patient alignments of the first three fractions. When patients begin feeling more comfortable and their positions become stable in the course of treatment, the possibility to contour an MVCT scan (in our case) and integrate it into the SIS for registration would improve handling significantly.

Compared with conventional marker alignment, the SIS improved positioning accuracy only for patients with \textit{pelvic} and \textit{chest} tumors. In these cases, the body surface to be captured is not covered, while patients with \textit{head \& neck} tumors wear a head mask. The mask keeps patients in the planning position. The patients have almost no degrees of freedom to misalign from planning position, being fully immobilized. The situation is different for patients with pelvic or chest cancers. Here, the risk for less accurate positioning is greater. The markers are arranged directly on the skin and can shift with the skin. The more or less unrestrained positioning of patients for irradiation of tumors in the \textit{Pelvic} and \textit{Chest} regions is a prerequisite for clinical use of SIS in radiation treatment,  especially when little use is made of other systems such as MeV or keV systems to monitor patient position.
%
%
\section{Conclusions}
\noindent Our comparison of the two SIS shows better performance of the AlignRT, which is likely attributable to the coverage of a larger body surface area in recording. The accuracy of both SIS differs between the three body regions investigated. The two systems were most accurate in the \textit{head \& neck} region, where conventional mask alignment is already sufficiently accurate. The other two body regions, \textit{pelvic} and \textit{chest}, might benefit from positioning by SIS, which are more accurate in these regions than conventional marker alignment. Nevertheless, there is still a need for improvement to reduce statistical variation. The \textsf{P$_{75}$} percentiles provide acceptable accuracies for clinical application for most distributions of both SIS, whereas the accuracies for the \textsf{P$_{95}$} (and \textsf{P$_{90}$}) percentiles are clearly above the planning safety margin.
%
%

\begin{backmatter}

\section*{Competing interests}
  The authors declare that they have no competing interests.

\section*{Author's contributions}
  LL conceived and designed the study. KK recorded the patient data and performed the statistical analysis. MW participated in the design of the study and wrote the article. All authors read and approved the final manuscript.

\section*{Acknowledgements}
  The authors thank the radiation therapists Katharina Berkovic, Jahed Abu-Jawad, and Christoph P\"{o}ttgen for adjusting patients from align to treatment positions using MVCT scans. Special thanks are due to the clinical staff operating the Tomotherapy treatment device during data acquisition. They independently handled the recording of the 3D surface with both SIS for almost every patient treated.

\section*{Support}
     This study was supported by Martin Stuschke, Director of the Radiation and Tumor Clinic at Universit\"{a}tsklinikum Essen.


\bibliographystyle{bmc-mathphys} 
\bibliography{References}      





%

\end{backmatter}
\end{document}